\documentclass[11pt]{artikel3}

\usepackage{amsmath,comment,graphicx,pslatex,verbatim,wrapfig}
\usepackage[pdftex,colorlinks]{hyperref}

\usepackage{acronym}
\acrodef{HPC}{High-Performance Computing}
\acrodef{KNL}{Knights Landing}

\def\n#{\bgroup \catcode`\_=12 \catcode`\>=12 \catcode`\<=12
  \catcode`\&=12 \catcode`\^=12 \catcode`\~=12 \def\\{\char`\\}\relax
  \tt \let\next=}

\title{Appearances of the Birthday Paradox in High Performance
  Computing}
\usepackage{authblk}

\author[1]{Victor Eijkhout\thanks{Corresponding author: \texttt{eijkhout@tacc.utexas.edu}}}
\author[2]{Margaret Myers}
\author[3]{John McCalpin}
\affil[1,3]{Texas Advanced Computing Center, The University of Texas at Austin}
\affil[2]{Department of Computer Sciece, The University of Texas at Austin}

\begin{document}
\maketitle

\begin{abstract}
  We give an elementary statistical analysis of two High Performance
  Computing issues, processor cache mapping and network port mapping.
  In both cases we find that, as in the birthday paradox, random
  assignment leads to more frequent coincidences than one expects
  \textsl{a~priori}. Since these correspond to contention for limited
  resources, this phenomenon has important consequences for performance.
\end{abstract}

\section{Introduction}

The birthday paradox is not so much a paradox as an unexpected
result. It expresses that collisions between unlikely events can be
more likely than expected \textsl{a prima facie}. 
In the phenomenon that gives it its name, we consider 
the probability that in a set of $n$ people a pair will
have the same birthday. The paradoxical aspect is that, for a given
probability, the value of $n$ is rather lower than one would
expect: for a $50\%$ chance it is enough to have $n=23$,
and $n=70$ gives a $99.9\%$ chance.

In this paper we give two examples of such collisions from the field
of \acf{HPC}, which is normally not associated with statistical
reasoning. 
We phrase the problem as follows: we consider random assignments of entities
to open slots; in cases where according to the `pigeonhole principle'
a conflict-free assignment is possible, in practice we will still get
conflicts. We then consider two common examples from the field of
\ac{HPC}, in particular cache associativity and network
switch oversubscription. In both these phenomena we show that
performance aspects can be given a simple statistical analysis.
We claim no particular novelty for this analysis, but we hope that our
exposition may provide both inspiring examples to statisticians and
an illustration of the use of statistics to \ac{HPC} practitioners.

References on the birthday problem:~\cite{Fisher:CarltonGalaBday,Mckinney:AMMbirthday}
For another application of `birthday' reasoning in computer science,
see~\cite{McGlynn:2001:BPL}.

\section{Appearance \#1: Memory caches}

\subsection{Background}

Since the late 1980's, the operating frequency of computer processors
has been increasing at a significantly faster rate than the operating
frequency of the computer's main memory.   To reduce processor stalls
while waiting on memory accesses, `caches' have been added to almost
all processors.   Caches are small amounts of high-speed memory that
transparently retain the data from recently accessed memory
locations. (They also retain the knowledge of where the data originated.)
Accessing a memory location that is mirrored in the
cache allows full-speed operation, while accessing a memory location
that does not have valid data in the cache typically results in lower
performance, since the
processor stalls waiting for data to arrive.

Since these high-speed caches are relatively small, the data from the
newly accessed memory location also displaces data from some
previously accessed memory location in the cache. This potentially
lowers performance again, since the displaced data may be needed in
the near future. Achieving high
performance thus requires algorithms that have high rates of data re-use on
subsets of the data that `fit' into the cache.

\subsection{Address mapping}

Our analysis of this data re-use, and hence code performance, will
focus on the \emph{mapping function} that translates memory addresses
into cache locations.
(Our discussion is a simplification, in that we
ignore the further memory access history of the application, and other 
non-trivial details of the cache
implementation.)

The mapping function is used every time the  processor performs a memory
access:  the memory address is translated in order to
determine which cache index to query. The cache control logic compares
the memory address requested by the processor with the memory address
held in that cache location.  If the addresses match, the data from
the cache is returned to the processor.  If the addresses do not
match, the request must be sent to the main memory to retrieve the
correct data.  When the data is returned from memory, the cache
replaces the previous data held at this index with the new data and
updates the memory address at this index with the address of the newly
loaded data.

\subsection{Direct mapping}

\begin{figure}[ht]
  \includegraphics[scale=.1]{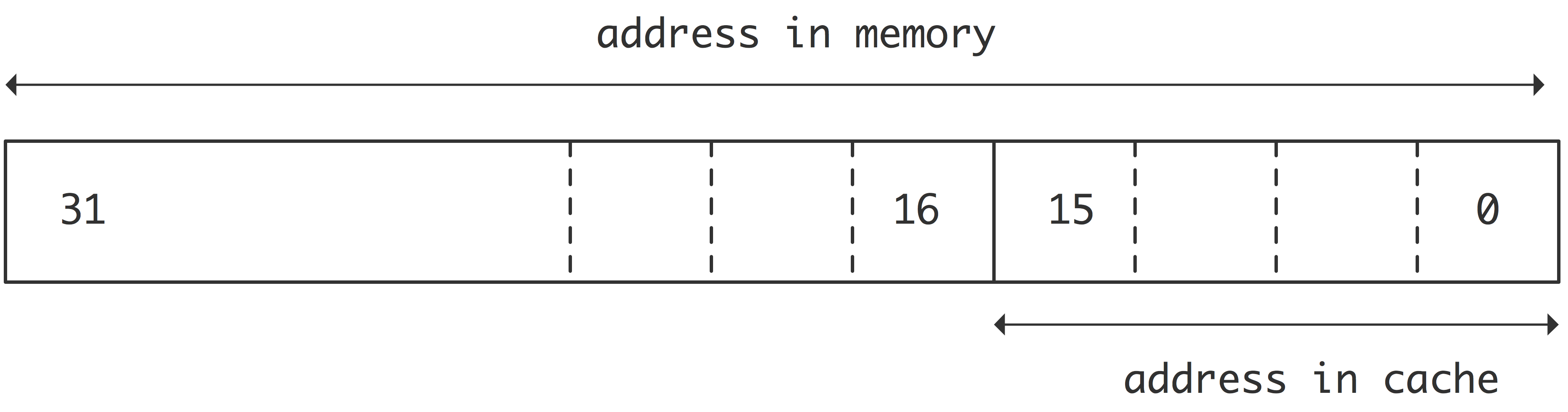}
  \caption{Direct mapping of 32-bit addresses into a 64K cache}
  \label{fig:directmap}
\end{figure}

From a hardware implementation perspective, the easiest way to map
memory locations to cache locations is with a power-of-2 modulo
function.   For a cache with $n=2^s$ locations, the cache `index'
consists of the first (low-order) $s$ bits of the memory address.
(Figure~\ref{fig:directmap} illustrates taking the last 16~bits of a
32~bit address.) At
each location, the cache stores a block of data, the corresponding
memory address, and a small amount of additional data to track
validity and other attributes.   

It is clear from this description that such a cache can hold $n$ data
blocks for re-use -- but only if the set of memory addresses being
accessed contains no duplicate cache index mappings.
In some cases, this can actually be guaranteed with
information available to the user.  For the direct-mapped cache
described above, any set of $n$ contiguous memory locations will map
to each cache index exactly once, guaranteeing freedom from conflicts
-- though it should be noted that user-visible addresses are subject
to a minimum of one level of address translation, which limits the
size of contiguous regions in machine-dependent ways.

\begin{figure}[ht]
  \includegraphics[scale=.1]{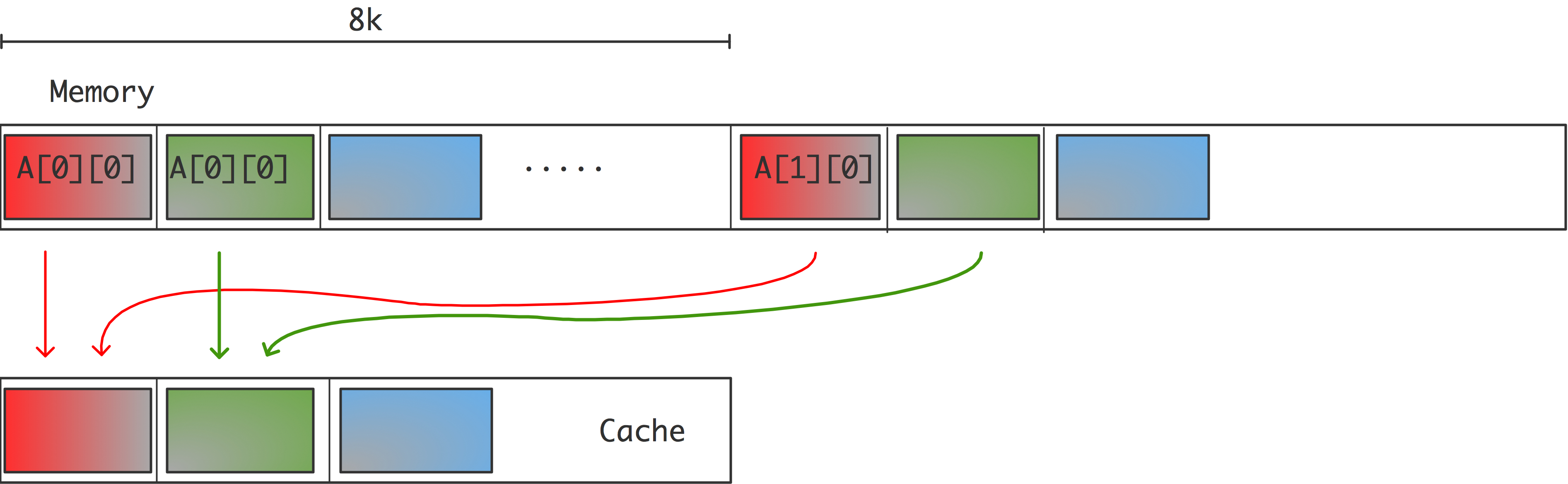}
  \caption{Mapping conflicts in direct mapped cache}
  \label{fig:directconflict}
\end{figure}

On the other hand, addresses at a distance of $n$ from each
other have the same $s$ lower address bits, and will then be mapped to
the same cache location. (This is illustrated in
figure~\ref{fig:directconflict}, which models the simultaneous
traversel of two matrix rows.)
Such a code will run much slower.

However, 
in many cases cache mappings are not visible to users in any
useful way, either because of virtual-to-physical address translations
(controlled by the operating system and not by the user), or because
the cache indexing function is complex and undocumented, or because
the set of addresses being accessed is dynamically randomized (e.g.,
with dynamically repartitioned graph structures).  In these cases, the
mapping of user-visible memory addresses to cache indices takes on a
random component, and statistical analysis becomes essential.

This brings us  to the `birthday problem'.   Given a
direct-mapped cache of size $n$ and a set of $p \leq n$ user-visible
memory addresses (that map to set of $p$ pseudo-random cache indices),
what can be derived about the probability of conflict of various
degrees?

As the number of sets grows, the probability ``zero conflicts''
approaches zero, but it also becomes less relevant.  A more relevant
metric might be the probability of getting a set of random addresses
with no more than some fixed percentage of conflicts -- maybe N/8 to
approach 90\% hit rates (but see the caveat above about how many cache
misses occur due to each conflict).  For a set of addresses that are
accessed only once per repetition and for which the order is the same
in each repetition, the number of hits is the number of sets with
exactly one cache line mapping --- all sets with 2 or more mappings
miss every time.

\textbf{Problem statement}
If we have a cache of $n$ locations, and we have a working set of $n$
memory words, some of these words will be mapped to the same cache
location, so we have an effective cache size that is smaller than $n$
elements. We will  investigate what the expected
effective cache size is.

\subsection{Associative caches}

\begin{figure}[ht]
\includegraphics[scale=.12]{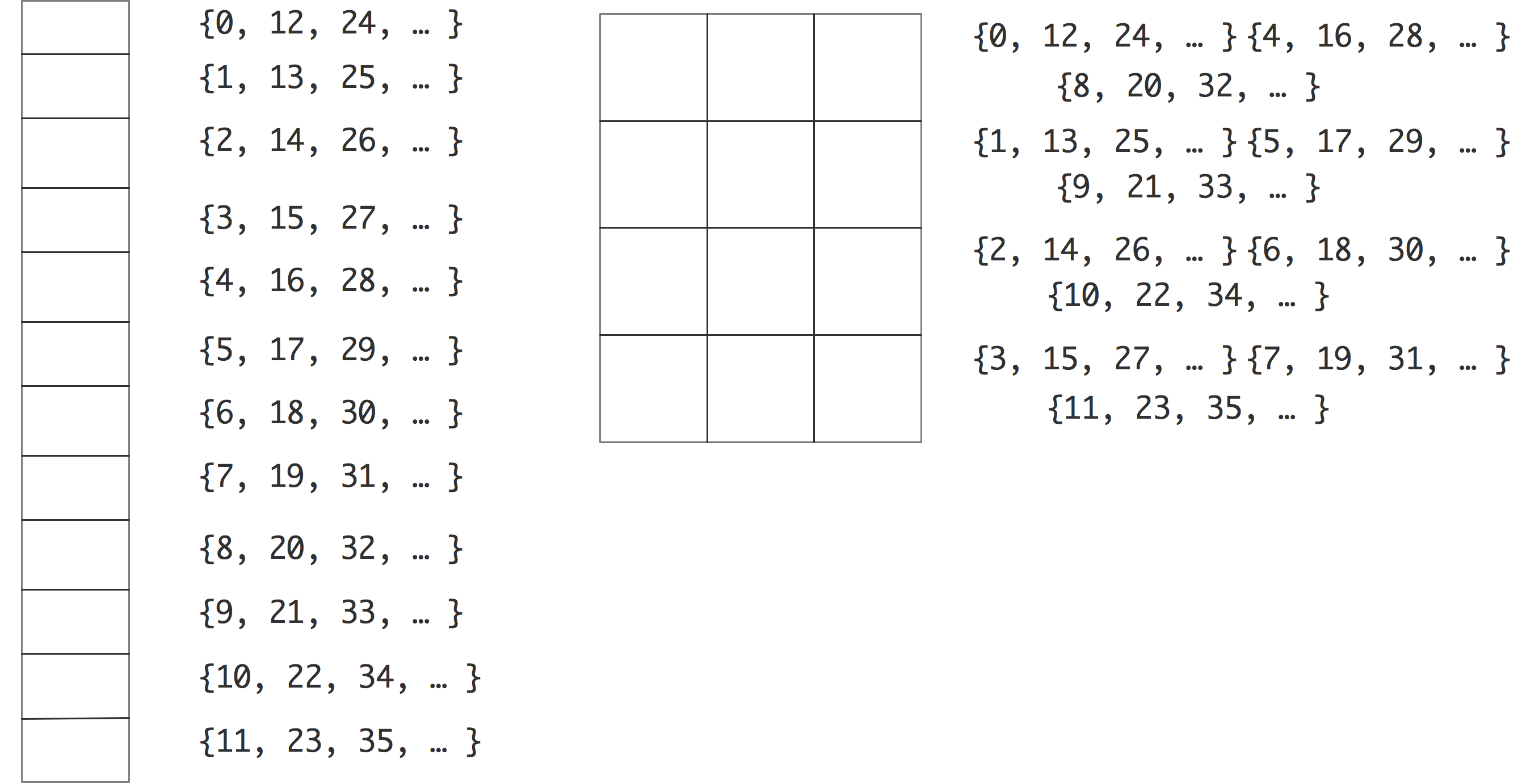}
\caption{Two caches of 12 elements: direct mapped (left) and 3-way associative (right)}
\label{fig:assoc-mapping}
\end{figure}

The common solution taken to mitigate mapping conflicts is to make
caches multi-way associative: addresses are no longer uniquely mapped,
but rather to a `set' of locations in the cache. We call a cache
$k$-way associative if each set has $k$ elements. This means that a
$k$-way collision can be resolved since $k$ identically mapped address
fill the $k$ locations of a set. This is illustrated in
figure~\ref{fig:assoc-mapping}. For instance, with~$k=4$, our code
can traverse one output and three input arrays that are at an otherwise
conflicting offset from each other.

\subsection{Model formulation}

We have just motivated the design of associative caches from traversing
multiple arrays in one algorithm. However, not all memory access is
this systematic. Thanks to out-of-order execution, multi-threading,
virtual-to-physical address translation, and any intrinsically
unstructured nature of the application,
we can also consider loads from memory as a random stream of
address requests. The question to analyze becomes then, what is the hit/miss
ratio of a random stream of addresses as a function of~$k$, the
associativity of the cache.

Increasing the value of $k$ is desirable since it
decreases the likelihood of conflicts. However, it also increases the cost,
both in design, construction, and energy usage of the electronics, as
well as possibly depressing the allowable clock speed. Our statistical
analysis is therefore part of the cost/benefit analysis of processor design.

We model our cache as follows: we assume a cache with $n$~locations, and
$k$-way associativity, giving $m=n/k$ sets.
An address~$a$ in main memory, is mapped to the set
$s(a)\mathop{:=}\mod(a,n/k)$, and each set can store $k$ items.
The question is then, given $n$ random addresses, what is the
likelihood that more than $k$ addresses~$a$ map to the
same~$s(a)$. This is a generalization of the `birthday paradox', which
in this context corresponds to $n=365$ and $k=1$.

\subsection{Analysis}

As with the regular birthday paradox, to analyze how likely collisions are, we
examine the complement: the probability of random mappings being
collision free. First we look at the expected number of addresses
stored if $n=mk$ addresses are mapped where $m$ is the number of sets, 
and we have $k$-way associativity.

Since the addresses are random, and all sets equally likely,
we use the linearity of
the expectation value to get our basic equation for the
expectation value of the number of items stored in
a cache with $k$ sets and $n=mk$ locations :
\[ E[ \#\mathrm{stored} ] = m E[ Y_i ], \qquad\hbox{$i\in\left[0,m\right)$}. \]
Our basic random variables are
\[ 
\begin{cases}
  X_i & \hbox{the number of addresses mapped to set~$i$}\\
  Y_i & \hbox{the number of addresses stored in set~$i$}
\end{cases}
\]
where we note that $Y_i=\min(X_i,k)$.

We derive $E[Y_i]$ by splitting it into two cases.
We observe that $Y_i$ has a maximum value of~$k$, the associativity,
but $X_i$ can be larger.  
If set~$i$
contains fewer than~$k$ elements, it means that exactly that many
addresses were mapped to it, and $Y_i=X_i$.
On the other hand,
if set~$i$ contains $k$ elements it means that $k$ or more addresses
were mapped to it, and the case $Y_i=k$
corresponds to the set of cases $X_i\geq k$.

This gives:
\newcommand\explain[1]{\noalign{\hskip\unitindent\small\textsl{#1}}}
\begin{align}
  E[Y_i] &= \sum_{j=0}^{k-1}jP(Y_i=j) + kP(Y_i=k) \\
  &= \sum_{j=0}^{k-1}jP(X_i=j) + kP(X_i\geq k) \\
\explain{take complement:}
  &= \sum_{j=0}^{k-1}jP(X_i=j) + k(1-P(X_i\leq k-1)) \\
\explain{enumerate cases $X_i\leq k-1$}
  &= k + \sum_{j=0}^{k-1}jP(X_i=j) + (-k)\sum_{j=0}^{k-1}P(X_i=j) \\
  &= k + \sum_{j=0}^{k-1} (j-k)P(X_i=j)\\
\explain{assuming $P(X_i=j)$ is binomial, with $n$ trials 
         and chance of success $1/m$}
  &= k + \sum_{j=0}^{k-1} (j-k) \Bigl({n \atop j}\Bigr) 
        \Bigl(\frac1m\Bigr)^j \Bigl( 1-\frac1m \Bigr)^{n-j}\\
\explain{simplify, using $m=n/k$:}
  &= k- \sum_{j=0}^{k-1} (k-j) \Bigl({n \atop j}\Bigr) 
        \Bigl( \frac{k}{n} \Bigr)^j \Bigl( \frac{n-k}{n} \Bigr)^{n-j}
        \\
        \label{eq:EY}
\end{align}

Example: for a cache of 1000 elements the number of elements stored
for a variety of associativities is:


\begin{tabular}{r|r|r|r|r|r|r|r|}
  \hline
  associativity&
  1&2&3&4&10&50&100\\
  \hline
  expected working set size&
  632&
  729&
  775&
  805&
  875&
  945&
  962\\  
  \hline
\end{tabular}

\textbf{Remark.} Since we essentially compute the expected number of
stored elements in a set, using a larger cache, while keeping
associativity, constant, only increases the number of sets. Therefore
the expected working set size is a fraction of the cache size that
depends on the associativity, not on the number of sets.
Under further assumptions on code behavior, such as a decaying `rate
of rereference~\cite{Hartstein:cache-sqrt}' we get a miss rate that
decays with cache size.

\subsection{Working set sizes}.

Now, if we want to prevent cache conflicts in other than a very
regular application we need to use a
working set that is smaller than the cache size. The above formula for
expected value is a good guideline, but what is the penalty for
getting conflicts regardless? For a single computer a slight loss of
efficiency is no emergency, for many processes
running a fairly tightly coupled application, this performance
degradation holds up the overall computation. 
Thus we could ask, given
a set of $P$ processors, what size working set can we safely
adopt to have a chance of conflicts~$<1/P$.
This is of course relevant for
clusters, which can have many thousands of processors, but we note that current
Intel Xeon processors can have up to 28 cores, making this question
already relevant on a single-computer scale.

We first analyze the chance of having no conflicts when $A$ random
addresses are assigned to a cache of $m$~slots and $k$-way
associativity, where~$A\leq\nobreak mk$. This is the number of
possible conflict-free mappings of $A$ addresses, divided by the total
number of mappings, which is~$m^A$.

Let $i_j$ be the number of
sets with $j$ addresses stored in them. While in principle all $i_j$ can
range~$0\ldots k$, there are also complicated relations between them,
giving recursively defined bounds. We say that $i_j$~can range from
$\min_j$ to~$\max_j$, which we will now compute. If $A_i$ is the number of
addresses mapped to slots with $\leq i$ addresses mapped to them, we have
$A_k=A$ and $A_{j-1}=A_j-j\cdot i_j$. With this we get\footnote
{The computation for $j=0,1$ is a special case as $i_0,i_1$ can take
  only one value: $i_1=\sum_{j>1}j\cdot i_j$ and $i_0=m-\sum_{j>0}i_j$.}
\[ \min_j=\max(0,A_j-(j-1)\sum_{j'>j}i_{j'}), \qquad \max_j = A_j/j. \]
Now for a given set of $i_j$ indices $i_0\cdots i_k$ we count the
number of permutations, giving us finally:
\begin{equation}
  P(\textrm{no conflict}) = \frac
  { \sum_{i_k=\min_k}^{\max_k}\cdots \sum_{i_0=\min_0}^{\max_0}
    \frac{m!}{i_0!\cdots i_k!} \cdot
    \frac{A!}{(0!)^{i_0}(1!)^{i_1}\cdots (i_k!)^{i_k}}
  }
  {m^A}
  \label{eq:pwork}
\end{equation}

If we apply this analysis, we see that a
probably-conflict-free working set will be quite small. For instance,
a cache with 4000 elements and 4-way associativity
will have an expected working set of around 3200 elements. However,
randomly assigning that many addresses will in fact have a probability
of only $0.025$ of no conflicts.


This table gives the probability of no-conflicts for various working
set sizes, in a 4-way associative cache of 4000 elements.


\begin{tabular}{r|r|r|r|r|r|}
  \hline
  working set size&
  $100$&  $200$&  $500$& $1000$& $2000$\\
  \hline
  no-conflict probability&
  $.999$& $.998$& $.843$& $2.31\cdot 10^{-2}$& $9.42\cdot 10^{-30}$\\
  \hline
\end{tabular}


Looking at this another way, we can ask if a certain fraction of
the cache size is a safe working set size. It turns out that for a
fixed fraction the
probability of no conflicts also goes down quickly with cache size.

\begin{tabular}{r|r|r|r|r|r|}
  \hline
  cache size&
  $100$&$200$&$500$&$1000$&$2000$\\
  \hline
  no-conflict probability&
  $.222$& $4.15\cdot 10^{-2}$& $2.73\cdot 10^{-4}$& $6.30\cdot 10^{-8}$&
  $3.34\cdot 10^{-15}$\\
  \hline
\end{tabular}

Having simultaneously low probability on hundreds or thousands of
processors will give an even lower number for the feasible working set size.

In this case we cannot but draw the conclusion that our analysis is overly
pessimistic, since it assumes totally randomly generated addresses.
In practice, programmers will use optimization techniques such as
blocking to make the memory use decidedly more regular.

\textbf{Computation remark} Equation~(\ref{eq:pwork}) takes some care
to compute. While both numerator and denominator are integers, and
languages such as python have support for very large integers, the
quotient is a real number, and the integers involved are too large to
convert to real. Hence all product have to be computed as powers of
sums of logarithms.

\subsection{Illustration from concrete examples}

\subsubsection{Modern Intel Xeon processors}
\label{sec:xeon}

In order to discuss a real-life cache we need to point out that
computer memory, as well as cache memory, is organized in
`cachelines', which are the smallest units of memory that can be moved
about. That is, if a program references a certain byte in memory, the
cacheline containing that byte is moved into cache. The typical
cacheline these days is 64 bytes, or 8 double precision numbers, long.

Now we can consider the L1 (data) cache of modern Intel Xeon processors,
which has a size of 32Kib, i.e., $2^{15}$~bytes. The typical
associativity is~8, meaning that 12~bits of an address are used to
determine the set. However, assignment is done per cacheline; not per
byte.
Since 6~of the address bits are used to determine the location of a
byte in the cacheline, the actual assignment uses the remaining (most
significant) 6~bits, meaning that the cache has 64 sets.

\[ \hbox{64 sets}\times \hbox{8-way associativity} \times
   \hbox{64 bytes per cacheline} = \hbox{$2^{15}$ bytes}.
\]

By the above analysis, we expect 441 out of 512 generated cachelines
to remain in the cache.

\subsubsection{Memory as cache in recent Intel processors}

In recent Intel processors, parts of the DDR4 memory can be configured
as direct mapped cache. The first architecture to do this was the
\ac{KNL} processor, but the same principle applies to the new `Intel
Optane DC Persistent Memory'.

\begin{figure}[ht]
  \includegraphics[scale=.6]{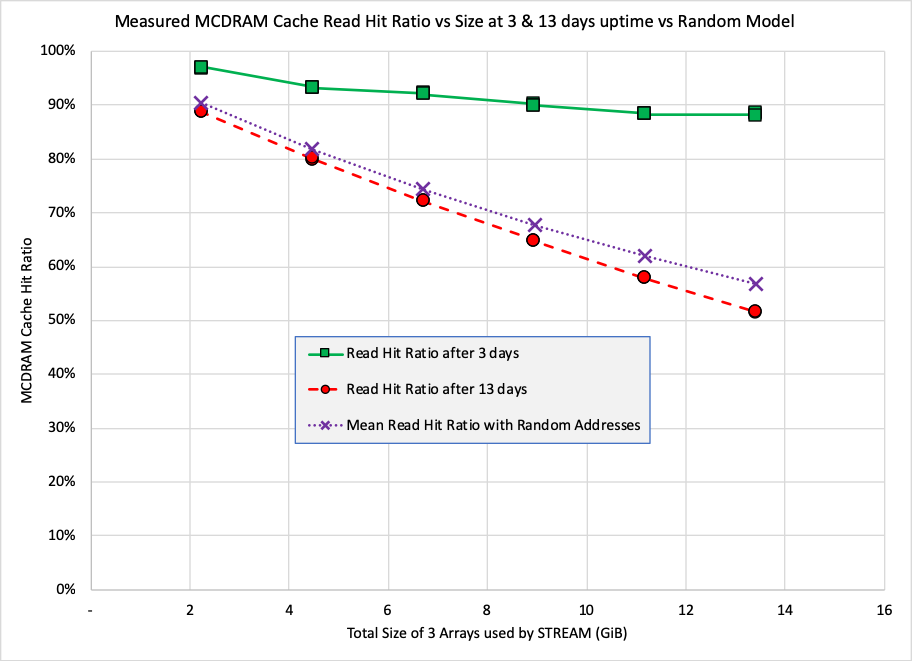}
  \caption{Performance of regular versus random memory access in a
    direct-mapped cache}
  \label{fig:knlstream}
\end{figure}

It was observed that codes using this cache mode had a decreasing
performance as as function of time. To understand this, it is
necessary to take into account the virtual-to-physical address
translation\footnote{Running programs refer to a memory address space
  expressed in virtual addresses; the page table translates to actual
  physical memory addresses.}, which is gradually established as programs run.
Initially, this translation fills up the cache linearly, so there are
no cache conflicts. Then, as the cache is full --~which can take a
while since this cache is typically 16Gbyte~-- the direct mapped
behavior becomes gradually apparent.

For instance, figure~\ref{fig:knlstream} shows how bandwidth (measured by the
Stream benchmark~\cite{STREAM}), which we use as a proxy for cache hit
rate, degrades as a
function of uptime of the processor: as time goes on the random mapping of pages turns
orderly memory access into random memory access.
In the context of supercomputer clusters, such as Stampede$\,$2 at the Texas
Advanced Computing Center, this was solved by reordering the list of
free pages after each user job.

\subsection{Discussion and related work}

We have shown that a model based on simple statistical tools can often
get close to observed behaviour. However, we had to make a number of
simplifying assumptions that make our story sometimes unduly
pessimistic. For practical processor design, therefore, one usually
relies on sophisticated modeling software, such as
CACTI~\cite{CACTI4}.

An early review paper on cache design~\cite{Smith:1982:CacheMemories}
used both simulations based on program traces, and gave a statistical
result on the likelihood of referencing data already in cache. The
latter uses the `LRU Stack Distance' model, which is 
based on the notion that, given a reference to a certain item an
memory, other items become more likely to be referenced next. The
resulting formula shows similarities to our result~(\ref{eq:EY}).

The classic computer architecture reference~\cite{HennessyPatterson} uses
simulation results from traces such as the Spec benchmarks. Unlike our
simplified model, it includes factors such as the time required for
looking up whether a data reference is a cache hit or miss.

The concepts behind caching were already
explored a few decades earlier in the context of
paging~\cite{McCo:pagedmemory}. Paging is usually presented as `moving
memory blocks --~or: pages~-- to disc when they are not in use, but
turning this story upside down, we see that it is essentially the same
phenomenon. A~program can use addresses out of a potentially large
logical address space, and pages get placed on physical addresses in
main memory. If a program uses a larger address space than there is
physical memory, or if multiple programs run simultaneously, memory
pages can be evicted to disc, which serves as bulk memory.

\section{Appearance \#2: Port mapping in supercomputer clusters}

\subsection{Background}

Supercomputer clusters typically contain hundreds to thousands of
`nodes', each containing a smal number of processor chips, typically
1 or 2 or~4. Since the cost of connecting each of $P$ nodes to every other
node is prohibitive, there is usually a network involved that limits
the number of connections to less than~$P^2$, while still offering
enough bandwidth for typical \ac{HPC} workloads. One popular network
topology is the `grid' where each node has a fixed, low number of
connections that arranges the nodes into the topological equivalent of
a grid. A~typical grid network is topologically three-dimensional, but
the K-computer has a 6-dimensional network.

\begin{figure}[ht]
  \begin{quote}
    \includegraphics[scale=.4]{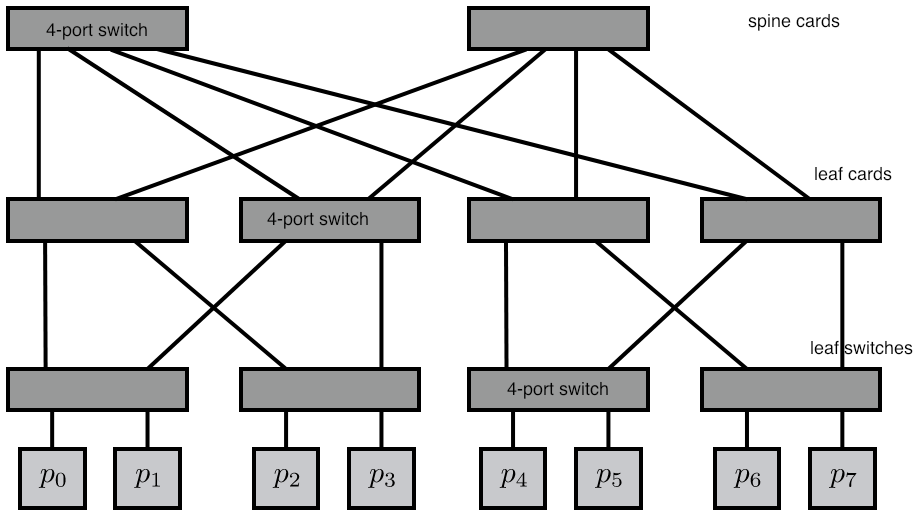}
  \end{quote}
  \caption{A multi-stage network for 8 processors built from 4-port switches}
  \label{fig:fattreeclos}
\end{figure}

The other currently popular topology is of a multi-stage network, built up out
of individual switching elements. This is known as a Banyan
network~\cite{Feng:interconnect}, a~`butterfly exchange'
network\footnote {\url{http://en.wikipedia.org/wiki/BBN_Butterfly}},
a~`fat-tree'~\cite{Leiserson:fattree} or a `Clos
network'~\cite{Clos1953}. Figure~\ref{fig:fattreeclos} depicts a
network where each switching element has four `ports', but in practice
a typical number is on the order of 30--40. The top level switches
have all their ports facing into the network, while the intermediate
ones designate half of their ports as `input' and half as `output' ports.

While mathematically the number of levels of such a network is a
logarithm of the number of processors, in practice the number of
levels is three, or in rare cases four. This is due to the high port
count, which functions as the base of the logarithm, and the existence
of multiple top level nodes.
\begin{figure}[ht]
  \includegraphics[scale=.15]{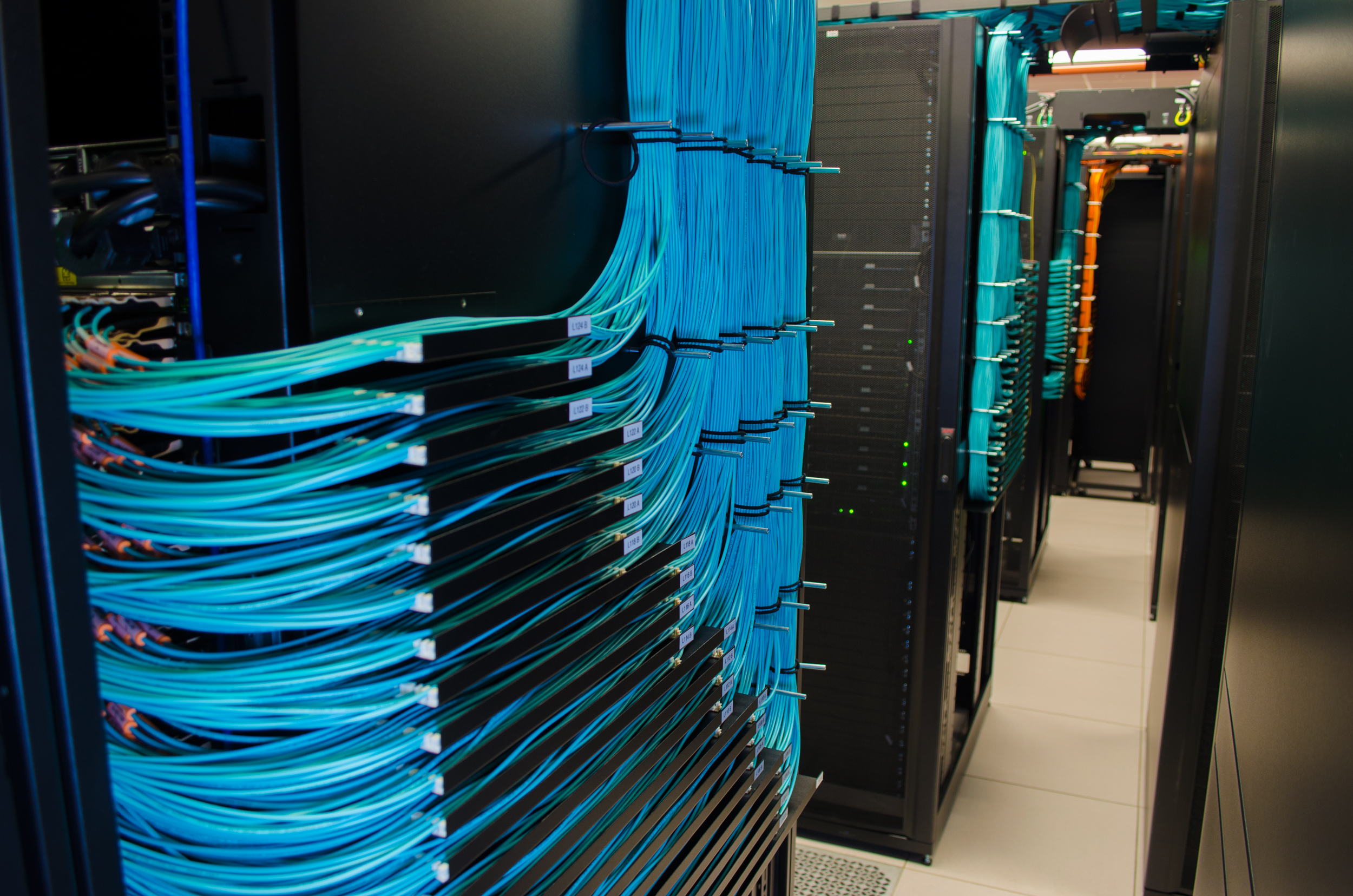}
  \caption{Networks switches for the TACC Stampede2 cluster (image
    courtesy Jorge Salazar, Texas Advanced Computing Center}
  \label{fig:taccswitches}
\end{figure}
As a practical illustration, figure~\ref{fig:taccswitches} shows three
(out of five) central switch cabinets of the Stampede2
cluster\footnote{\url{https://www.tacc.utexas.edu/systems/stampede2}},
recognizable by the blue cabling. These cabinets contain both the
top-level `spine cards', as well as the intermediate level `leaf
cards'. The `leaf switches' connected to the processors are part of
the processor cabinets.

\subsection{Problem statement}

\begin{figure}[ht]
  \begin{quote}
    \includegraphics[scale=.4]{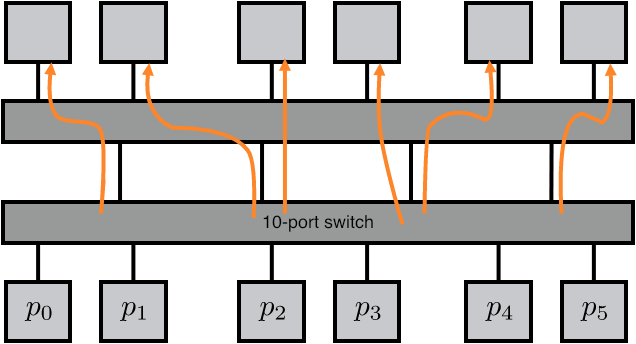}
  \end{quote}
  \caption{Over-subscription on a 10-port switch with routing as
    function of destination indicated.}
  \label{fig:oversubscription}
\end{figure}

The design of the network is not uniquely determined by the number of
processors and the number of ports on a node.
At the top level of the network (confusingly, also known as the root
nodes) all ports are connected to the next, intermediate,
network layer. However, at the intermediate layers there is the possibility
to trade `outbound' ports for `inbound' ones, only keeping the total
number of ports constant. This process is known as
`oversubscription', since it means that there can be more traffic
going into the switch than is going out.
Oversubscription is an obvious cost saving measure, since it
cuts down on the number of switching elements needed at higher levels.

\begin{figure}[ht]
  \begin{quote}
    \includegraphics[scale=.4]{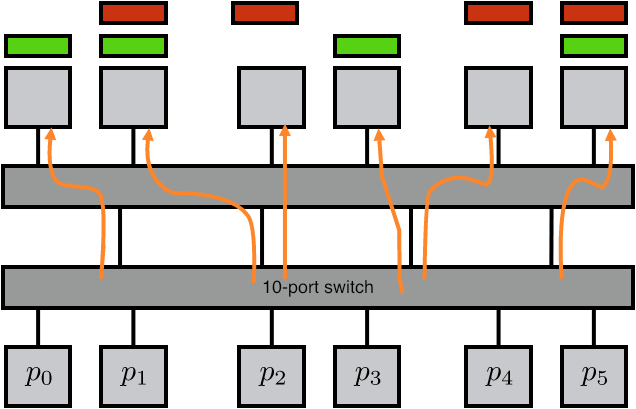}
  \end{quote}
  \caption{Examples of destination processors that are reached without contention
    (green) and with (red).}
  \label{fig:contention}
\end{figure}

However, oversubscription means that we have to map $n$ incoming
connections to $k<n$ output ports, and so there is obviously
contention if all $n$ nodes are sending at the same time. 
Figure~\ref{fig:oversubscription} shows a simplified
network where we have done away with all intermediate stages and only
show sending and receiving processors plus the switches they are
directly attached to. In this figure we route
traffic for six processors through four outbound ports on a 10-port
switch; the arrows show the mapping from target processor to output
port.

Now consider that, if $k$ nodes are sending through the switch, we have
theoretically enough capacity, but the `static routing' used by most
clusters means that some choices of $k$ destination nodes
will give contention for the output
ports. Figure~\ref{fig:contention} shows two sets of destination
processors; the green set has no contention, while the red set does.
Here we have another case where according to the pigeon-hole principle
a unique assignment should be possible, but where in practice
conflicts are likely.

This is then our problem: to analyze the chance of collision-free
traffic if the number of simultaneous messages equals that of the
number of output ports. Having simultaneous messages is an expected
scenario in practice, since scientific application often behave in a
synchronized manner, even if they do not have explicit synchronization
mechanisms.

\subsection{Analysis}

In the traditional birthday problem, we have a large space of
possibilities and a small number of picks from that space. Here we
have a much smaller space of possibilities, and the number of picks is
equal to it. We could model this as follows: consider birth
\emph{months} rather than days, and we have more than 12
people. Picking 12 out of these, what is the chance that no two share
a birth month?

\begin{figure}[ht]
  \includegraphics[scale=.8]{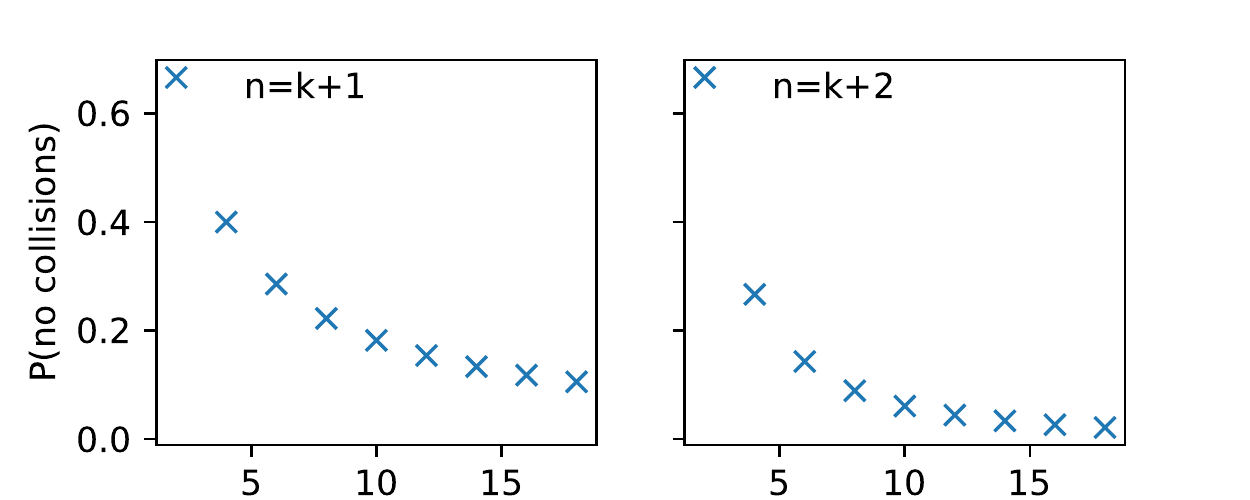}
  \caption{The probability of no message collisions for one-way ($n=k+1$,
    left) and
    two-way ($n=k+2$, right)  oversubscription as function of the
    number of destinations~$k$}
  \label{fig:portcollide}
\end{figure}

Let $n$ be the number of destination nodes, $k$~the number of ports,
and $k$~be the number of requests.  Each port has 1 or 2 associated
destinations, so ports will have 0, 1 or 2 requests.  If a port has 2
requests, that is a collision.  Let $C$ be the number of
collisions. We will categorize ports as either single or double,
depending on whether one or two messages target them.
There will be $n-k$ double ports and $2k-n$ single.  If order of
destination requests is not considered, there are $C(n,k)$ ways of
choosing $k$ destinations of the $n$ without other restrictions.  The
number of ways of making $k$ requests without having collisions is
found by counting the number of ways that each port has exactly one
request. With order not considered, there is one destination choice
for each single and 2 for each double port, This yields $2^{n-k}$ unique
port configurations, so the probability of no collisions
is~$2^{n-k}/C(n,k)$.

An interesting way to get insight in this formula is to use
\[ \frac{P(\hbox{\scriptsize No collisions for $n$})}
   {P(\hbox{\scriptsize No collisions for $n-1$})} = \frac{2(n-k)}{n}.
\]
Using that $P=1$ for $n=k$, we get (for $k\geq 2$):
\[ 
\begin{array}{rl}
  \hbox{if $n=k+1$:}&{P(\hbox{\scriptsize no collisions})}=\frac2{k+1}\\
  \hbox{if $n=k+2$:}&{P(\hbox{\scriptsize no collisions})}=\frac8{(k+1)(k+2)}\\
\end{array}
\]
The case $n=k+1$ is not practical, since the number of ports
$p\equiv k+n$ on a switch is typically even. That leaves the second
case as the minimal one, and in terms of~$p$:
\[ 
  n=k+2\colon P(\hbox{\scriptsize no collisions})=\frac8{n(n-1)}=\frac{32}{p(p+2)}.
\]
In a 30-port design, a quite reasonable size, the minimal
oversubscription is then 16~input ports and 14~output ports.
In a random communication with 14 sources and
destinations this gives a
chance of no collisions  of~$1/30$.

\subsection{More discussion}

In the above discussion we made it sound as if a message chooses a
port. In practice, the port is `destination routed': the port is
statically determined by the message destination according to the
`routing tables'.
These map, on each switching element, the possible destinations to the
output port on that switch. While it would be possible to configure a
program run so as to take advantage of the routine tables, in practice
this is infeasible, the more so since conflicts with other users'
programs are upredictable.

This particular instance of the birthday
paradox is alleviated with dynamic routing, 
which will be available on the TACC Frontera machine, to come online
medio 2019.

\bibliography{vle}
\bibliographystyle{plain}

\end{document}